\documentclass{article}

\usepackage{spconf,amsmath,graphicx,hyperref}
\usepackage{booktabs,threeparttable,makecell}
\usepackage{xcolor,amssymb,multirow}
\usepackage{booktabs,tabularx}
\usepackage{booktabs}
\newcommand{\valci}[3]{#1\,(\,#2--#3\,)}

\usepackage{amsmath,mathtools}
\newcommand{\Yes}{{\color{ForestGreen}\checkmark}}
\newcommand{\No}{{\color{BrickRed}$\times$}}

\usepackage{booktabs}
\usepackage{makecell}
\usepackage{threeparttable} % threeparttable + tablenotes
\usepackage{adjustbox}      % 等比例缩放整块内容
\usepackage{pifont}         % \ding 图标
\usepackage{caption}        % 更稳健的 caption 处理（可选）
\usepackage{hyperref}
\usepackage{fontawesome5} % 提供 \faGithub
\hypersetup{colorlinks=true, urlcolor=blue}
\usepackage[dvipsnames]{xcolor}   % for ForestGreen/BrickRed
\usepackage{pifont}               % for \ding check/cross symbols

\title{MMedFD: A Real-world Healthcare Benchmark for Multi-turn Full-Duplex Automatic Speech Recognition}
\name{%
  \shortstack[c]{%
    Hongzhao Chen\textsuperscript{*1}\thanks{*Co-first author},
    XiaoYang Wang\textsuperscript{*2},
    Jing Lan\textsuperscript{*1},
    Hexiao Ding\textsuperscript{1},
    Yufeng Jiang\textsuperscript{1},\\
    \textit{MingHui Yang\textsuperscript{2}, DanHui Xu\textsuperscript{2}, Jun Luo\textsuperscript{2},
    Nga-Chun Ng\textsuperscript{1}, Gerald W.Y.\ Cheng\textsuperscript{1},
    Yunlin Mao\textsuperscript{1}, Jung Sun Yoo\textsuperscript{\#1}}%
  }%
  \thanks{\#Corresponding author}
}
\address{\textsuperscript{1}Department of Health Technology and Informatics\\The Hong Kong Polytechnic University, Hong Kong SAR, China\\
    \textsuperscript{2}  Multimedia and Algorithm Quality Team \\
   AntGroup, HangZhou, China\\
    Emails: \{hongzhao.chen, jing-hti.lan, hexiao.ding, yufeng.jiang, yunlin.mao\}@connect.polyu.hk\\
 \{wai-yeung.cheng, jungsun.yoo\}@polyu.edu.hk\\
 \{sam.nc.ng\}@hksh.com, \{yangxiao.wxy, minghui.ymh, xudanhui, wuguo.lj\}@antgroup.com\\
  \small
  \faGithub\ \href{https://github.com/Kinetics-JOJO/MMedFD}{\texttt{https://github.com/Kinetics-JOJO/MMedFD}}\\
  \small
  \faDatabase\ \href{https://huggingface.co/datasets/HanselZz/MMedFD}{\texttt{https://huggingface.co/datasets/HanselZz/MMedFD}}
}

\begin{document}
%\ninept
%
\maketitle

\begin{abstract} 
Automatic speech recognition (ASR) in clinical dialogue demands robustness to full-duplex interaction, speaker overlap, and low-latency constraints, yet open benchmarks remain scarce. We present MMedFD, the first real-world Chinese healthcare ASR corpus designed for multi-turn, full-duplex settings. Captured from a deployed AI assistant, the dataset comprises 5,805 annotated sessions with synchronized user and mixed-channel views, RTTM/CTM timing, and role labels. We introduce a model-agnostic pipeline for streaming segmentation, speaker attribution, and dialogue memory, and fine-tune Whisper-small on role-concatenated audio for long-context recognition. ASR evaluation includes WER, CER, and HC-WER, which measures concept-level accuracy across healthcare settings. LLM-generated responses are assessed using rubric-based and pairwise protocols. MMedFD establishes a reproducible framework for benchmarking streaming ASR and end-to-end duplex agents in healthcare deployment. The dataset and related resources are publicly available at \url{https://github.com/Kinetics-JOJO/MMedFD}.
\end{abstract}


\section{Introduction}
\label{sec:intro}

Automatic speech recognition (ASR) transcribes speech into text and is increasingly used in clinical workflows \cite{adedeji2024sound}. High-quality transcripts improve documentation and enable downstream tasks such as entity capture, concept linking, and note generation for patient care and auditing \cite{shi2024medical,leduc2024multimed}. ASR remains challenged by real-world healthcare dialogue with multi-turn and highly dynamic content. These conversations include frequent interruptions and overlapping speech and require reliable speaker diarization. Streaming systems also need robust context tracking, recovery after barge-in, and low latency \cite{valizadeh2022doctor}.

Recent full-duplex dialogue systems coordinate listening and speaking via learned controllers and substantially reduce latency relative to half-duplex interaction \cite{wang2024fullduplex}. This motivates integrating streaming ASR with overlap-aware control for deployment. Chinese clinical ASR still lacks open corpora and harmonized evaluation standards, and public real-world benchmarks remain scarce. PriMock57 \cite{papadopoulos-korfiatis-etal-2022-primock57} provides simulated primary care consultations for controlled studies, but it also underscores the shortage of publicly available real-world Chinese clinical ASR corpora with standardized splits.

Multi-turn spoken benchmarks increasingly use rubric-guided evaluation to assess competence beyond literal word accuracy \cite{du2025mtalkbench}. Healthcare dialogue has similar needs, where word error rate (WER) alone does not capture medical concepts, dialogue coherence, or role consistency. Post-processing with large language models (LLMs) can reduce WER and improve medical concept recognition, but its effectiveness under multi-turn full-duplex conditions is not established \cite{adedeji2024sound}. Recent surveys highlight this gap and motivate end-to-end architectures that jointly model recognition, speaker diarization, and dialogue management under operational constraints \cite{leduc2024multimed,valizadeh2022doctor}.

\begin{figure*}[!t]
  \centering
  \includegraphics[width=\textwidth]{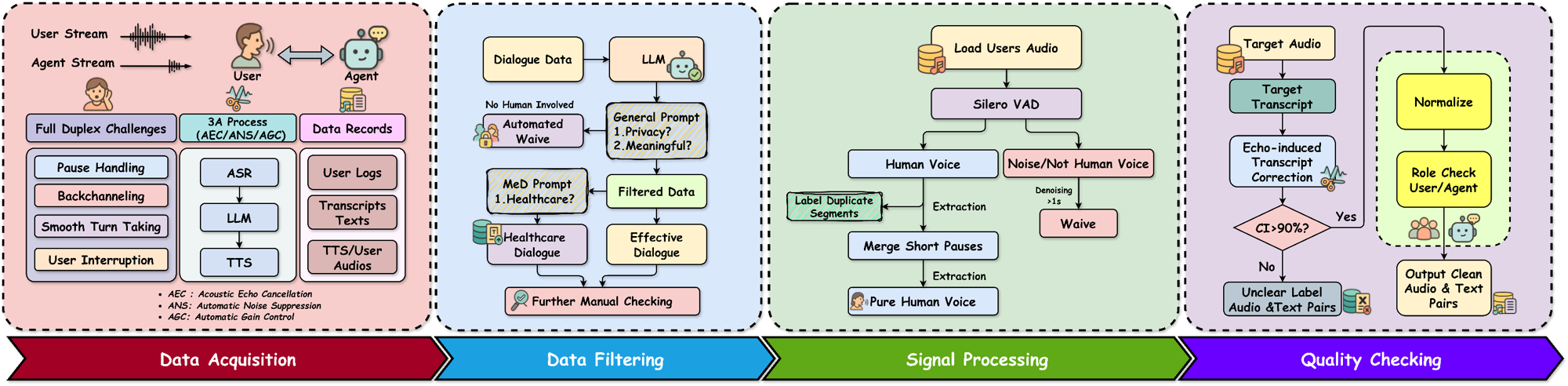}
    \caption{Overview of the four-stage pipeline from left to right, including Data Acquisition, Data Filtering, Signal Processing, and Quality Checking, producing clean, high-quality audio and text pairs.}
  \label{fig:benchmark-overview}
\end{figure*}

We introduce MMedFD, a real-world benchmark for multi-turn full-duplex Chinese healthcare dialogue. Figure~\ref{fig:benchmark-overview} summarizes the protocol for corpus construction and dialogue evaluation. We make three contributions.

\begin{enumerate}
\item We release a real-world Chinese healthcare dialogue corpus with multi-turn structure, collected with consent and de-identified, annotated with role labels and medical entities for supervised training and evaluation.
\item We present a model-agnostic full-duplex pipeline that standardizes streaming segmentation, speaker diarization, context packaging, and cross-turn memory to support real-time barge-in and overlap handling.
\item We evaluate ASR with WER, Healthcare Concept WER (HC-WER), entity-level precision and recall, and metrics for dialogue coherence and role consistency.
\end{enumerate}

% ################version 4
% \begin{figure}[t]
%   \centering
%   \includegraphics[width=\textwidth]{section/img/data_pl.png} % 路径按你的工程调整
%   \caption{Overview of benchmark construction.}
%   \label{fig:benchmark-overview}
% \end{figure}
% 文中
\section{Related Work}
Traditional spoken dialogue systems mostly rely on half-duplex paradigms that execute a rigid, sequential turn-taking protocol. To deliver fluid interaction, full-duplex systems have recently gained substantial traction~\cite{5, 7}. This shift requires the assistant to concurrently process streaming user inputs and manage proactive conversational behaviors including interruption handling, backchanneling, and smooth turn-taking during continuous audio playback~\cite{5}.

However, full-duplex operations introduce severe upstream acoustic complexities. The concurrent execution of text-to-speech synthesis and user barge-in causes massive acoustic echo and non-stationary ambient noise, demanding robust acoustic echo cancellation and noise suppression frameworks~\cite{8}. Residual echo and spontaneous speaking styles also heavily degrade downstream speaker diarization and context tracking~\cite{12}. While benchmarks like SpokenWOZ~\cite{19} and VoxDialogue~\cite{20} evaluate multi-turn conversational agents in general domains, they overlook the strict medical concept fidelity and role-consistency constraints required in clinical consultations. MMedFD bridges this critical gap by providing the first real-world full-duplex healthcare benchmark.

\section{Methodology}
\label{sec:method}
\subsection{Data Acquisition}
\label{subsec:data-acq}
We collected speech from a full-duplex healthcare assistant (agent) during internal beta testing. The agent continuously listened while generating synthesized replies. To match deployment acoustics, we retained acoustic echo cancellation (AEC), automatic noise suppression (ANS), and automatic gain control (AGC) \cite{aec_eval}. Each session was stored as long-form single-channel 16 kHz 16-bit pulse code modulation (PCM) audio with wall-clock timestamps, enabling Rich Transcription Time Marked (RTTM) turn annotation and Conversation Time Marked (CTM) word alignment \cite{nist_rttm}. For each agent reply, we stored the text produced by the LLM, a text-to-speech (TTS) waveform, a stable reply identifier and order, role labels, and a session manifest linking text and audio. We exported two synchronized views without filtering. The conversation view retained the complete mixed timeline. The user view excluded non-user segments while preserving residual echo to reflect full-duplex artifacts.

\subsection{Data Preprocessing}
\label{subsec:preprocess}
Preprocessing combined dialogue filtering with signal conditioning. An LLM screened for personally identifiable information (PII), semantic validity, and healthcare relevance, following governance-first curation practices consistent with the BigScience ROOTS corpus \cite{laurencon2022roots}. PII-positive items were blocked from annotator access and removed. Remaining content was manually verified and redacted before release. Audio conditioning included DC offset removal, peak limiting, root mean square (RMS) normalization, and band-limiting. Segmentation used Silero voice activity detection (VAD) \cite{silero_vad} with padding and short-gap merging. Non-speech audio longer than one second was excluded from the user view and retained in the conversation view. Playback-aware diarization based on pyannote-style modeling refined user–agent boundaries and identified overlapping speech excluded from supervised training \cite{pyannote}. Residual echo was detected by correlating the microphone signal with the agent TTS reference and was masked or down-weighted \cite{aec_eval}. This stage produced RTTM turns, segmentation for the user view with per-segment signal-to-noise ratio (SNR), and updated session manifests.

\subsection{Quality Control}
\label{subsec:quality}
Quality control verified linguistic fidelity, privacy, role labeling, and split hygiene. A domain-adapted end-to-end ASR model transcribed the user view. We applied forced alignment on the conversation view to obtain word timings and produced CTM files consistent with RTTM \cite{ctm_sctk,nist_rttm}. Annotators followed predefined rules. TTS leakage caused by AEC failures was not considered user speech. Segments with acoustic ambiguity or ASR confidence below 0.90 were marked unclear. Text normalization handled numerals, dates, units, drug names, and abbreviations while preserving alignment with audio. This preserved semantic-acoustic correspondence and improved data quality. De-identification replaced personal identifiers with placeholders. Dual validation used human annotators and an LLM. We enforced session-level splits to avoid leakage when user and agent spoke concurrently. Each release included raw conversation audio, user segments, RTTM and CTM files, normalized transcripts with placeholders, and a manifest of known artifacts.

\section{Experiments}
\label{sec:exp}

% \subsection{Experimental Setup}
% We fine-tuned Whisper-small \cite{radford2023whisper} end-to-end on the Chinese training split for ASR in healthcare dialogue. We constructed role-concatenated audio streams by merging all $User$ turns and all $Agent$ turns separately. This supports long-context normalization and improves coverage of clinical terminology without turn-level alignment. Models were trained on Chinese only and evaluated on the held-out Chinese test split.
% \subsection{Experimental Setup}
% We fine-tuned the Small, Medium, and Large architectures of the Whisper model \cite{radford2023whisper} end-to-end utilizing the Chinese training split for healthcare dialogue ASR. By independently concatenating all \textit{User} turns and \textit{Agent} turns, we generated role-specific audio streams. This approach leverages long-context normalization to improve clinical terminology coverage without the need for strict turn-level alignment. The models were trained solely on Chinese data and assessed on a held-out Chinese test split.
\subsection{Experimental Setup}

Whisper \cite{radford2023whisper} is a robust foundation model for speech recognition, pre-trained on a vast corpus of diverse multilingual audio. While it exhibits strong zero-shot capabilities in general domains, its performance often degrades when applied to specialized healthcare dialogues. This domain gap primarily stems from the high density of complex clinical terminology, medication names, and the conversational nuances inherent in spoken Chinese medical consultations. To mitigate this discrepancy and adapt the foundation model to our specific clinical domain, domain-specific fine-tuning is essential.

To investigate the impact of model capacity on transcription accuracy, we selected three distinct architectures of the Whisper model: Small, Medium, and Large. The architectural specifications and parameter counts for these variants are detailed in Table~\ref{tab:whisper_params}. 

% ==================== 双栏一侧专用表格 (4列极限压缩版) ====================
\begin{table}[t] 
  \centering
  \footnotesize % 必须使用 \footnotesize 或 \scriptsize 来适配 4 列
  \caption{Parameter specifications and language support of the evaluated Whisper variants.}
  \label{tab:whisper_params}
  
  % 将列间距压缩到 4pt，这是双栏排版塞下 4 列的极限安全距离
  \setlength{\tabcolsep}{4pt} 
  \renewcommand{\arraystretch}{1.15}
  
  \begin{tabular}{@{} l c c c @{}}
    \toprule
    \textbf{Model} & \textbf{Parameters} & \textbf{English-only} & \textbf{Multilingual} \\
    \midrule
    Whisper-small  & 244 M  & \checkmark & \checkmark \\
    Whisper-medium & 769 M  & \checkmark & \checkmark \\
    Whisper-large  & 1550 M & $\times$   & \checkmark \\
    \bottomrule
  \end{tabular}
\end{table}
% ======================================================================
We fine-tuned these models end-to-end utilizing the Chinese training split for healthcare dialogue ASR. Rather than processing isolated short utterances, we generated role-specific audio streams by independently concatenating all \textit{User} turns and \textit{Agent} turns. This formulation leverages the inherent long-context capabilities of the Whisper architecture, enabling better cross-utterance normalization and enhancing the capture of clinical terminology without necessitating strict turn-level alignment. All models were trained solely on the Chinese dataset to prevent catastrophic forgetting of the target language and evaluated on a held-out Chinese test split.
\subsection{Implementation Details}
For all Whisper variants, we applied the same data processing and training setup. To ensure clean text targets, we normalized the transcripts by standardizing punctuation, numerals, dates, measurement units, drug names, and medical abbreviations, all while maintaining strict alignment with the audio. We trained the models on 4$\times$ NVIDIA L20 GPUs for 1{,}000 epochs, using a batch size of 8 and a learning rate of $1\times 10^{-4}$. Decoding hyperparameters for both ASR and response generation were tuned on the development set and kept fixed during testing.

% % \subsection{Speaker Diarization Assessment}
% % Accurate role attribution is vital for medical documentation. We evaluated the speaker diarization pipeline (pyannote-audio) using Diarization Error Rate (DER) and Jaccard Error Rate (JER) on the full session audio.
% % The system achieved a DER of \textbf{[XX.X]\%}. Most errors stemmed from short backchannels ($<0.5$s) in the \textit{Dual-Overlap} subset, suggesting that future work must jointly model ASR and diarization to resolve rapid turn-taking in clinical contexts.
\section{Results}
\label{sec:res}

\subsection{Benchmark Description}

MMedFD is a benchmark for Chinese healthcare spoken dialogue constructed from live user–agent interactions under full-duplex operation. The corpus contains 136.9 hours of role-labeled audio and turn-segmented transcripts spanning 1,805 dialogues and 10,814 turns. As shown in Table~\ref{tab:mt-fd-hc}, most real-world healthcare corpora lack explicit multi-turn annotation and rarely include full-duplex recordings. To our knowledge, MMedFD is the first Chinese healthcare benchmark collected in real-world settings and annotated at multi-turn granularity under full-duplex operation. It supports systematic evaluation of duplex dialogue systems and ASR models under realistic timing constraints, including interruption handling and turn-taking behavior in healthcare scenarios.

\subsection{Evaluation Results}
We report two studies. The first evaluates ASR across Whisper model scales, including \textit{whisper-small}, \textit{whisper-medium}, and \textit{whisper-largev3}, using role-specific metrics. The second evaluates downstream healthcare response quality by feeding transcripts from the best ASR configuration into multiple text-only LLMs.

\subsubsection{Role-wise ASR Performance and Scaling Behavior}
Table~\ref{tab:role_metrics_models} reports ASR metrics on the train and test splits. User speech is consistently harder than Agent speech. For \textit{whisper-medium}, Test WER is 35.25\% for User and 13.04\% for Agent. This gap reflects uncontrolled patient-side acoustics and spontaneous speaking styles, while Agent speech is cleaner and more regular.

Scaling improves robustness on User speech. Test WER drops from 53.11\% with \textit{whisper-small} to 12.91\% with \textit{whisper-largev3}. In contrast, Agent metrics do not improve with scale. \textit{whisper-small} achieves the lowest Agent WER at 1.84\%, and larger variants regress on this subset. Healthcare Concept WER (HC-WER) further shows that clinical terminology remains a bottleneck. Despite better overall WER, \textit{whisper-largev3} yields a User Test HC-WER of 36.84\%, higher than smaller variants. This suggests that scaling alone does not guarantee concept-level precision without targeted domain adaptation.

\subsubsection{Downstream LLM Response Quality}
We evaluated downstream response quality with fixed ASR transcripts and compared our pipeline with proprietary and open-source text-only large language models (LLMs) in Table~\ref{tab:stacked_paireval_geval}. Under PairEval, our pipeline is competitive and reaches a 49.0\% Win rate against strong baselines. Tie outcomes dominate across model pairs, with Tie rates between 95.0\% and 99.8\%. This suggests that for common healthcare queries, pairwise preferences are weak and models often produce comparable answers.

G-Eval shows a similar pattern. Our pipeline achieves an Overall score of 4.0 out of 5, comparable to Claude-Opus-4.1 at 4.1 and Gemini-2.5-Pro at 4.0. Correctness and Safety are 36.9\%, supporting that the generated responses follow clinical facts and safety constraints. These results indicate that domain-specific tuning can match general-purpose models on clinical dialogue response quality.

\begin{table*}[!htbp]
\centering
% 用 adjustbox 将 threeparttable（含表+注释）整体缩放到 \textwidth
\begin{adjustbox}{width=\textwidth}
\begin{threeparttable}
\caption{Comparison of spoken-dialogue benchmarks and their evaluation protocols, including healthcare setting, scale, and language.}

\label{tab:mt-fd-hc}

\renewcommand{\arraystretch}{0.92}
\setlength{\tabcolsep}{1.8pt} % tighter column spacing

\begin{tabular}{l | c c c | c | c c c c | c | c}
\toprule
\textbf{Benchmark} &
\makecell{\textbf{Multi-}\\\textbf{Turn}} &
\makecell{\textbf{Full-}\\\textbf{Duplex}} &
\makecell{\textbf{Health-}\\\textbf{Care}} &
\textbf{Nature} &
\textbf{Roles} &
\textbf{Dur.\,(h)} &
\textbf{\#Dialogs} &
\textbf{\#Turns} &
\textbf{Language} &
\multicolumn{1}{c}{\textbf{Evaluation}}\\
\midrule
MultiMed\cite{leduc2024multimed}                       & \No  & \No  & \Yes & Real-world & 6  & 150      & --     & --       & Multiling.       & ASR(WER/CER) \\
VietMed\cite{le-duc-2024-vietmed}                      & \No  & \No  & \Yes & Real-world & 6  & 16       & --     & --       & Vietnamese       & ASR(WER) \\
PriMock57\cite{papadopoulos-korfiatis-etal-2022-primock57} & \Yes & \No  & \Yes & Simulated  & 2  & 9        & 57     & $\approx$5244    & English         & ASR(WER), Task \\
Fareez et al.\cite{fareez2022respiratory}              & \Yes & \No  & \Yes & Simulated  & 2  & 55.15    & 272    & --       & English         & ASR(WER), Human-Eval \\
myMediCon\cite{htun-etal-2024-mymedicon}               & \No  & \No  & \Yes & Read       & 2  & 11       & --     & --       & Burmese         & ASR(WER) \\
AfriSpeech-200\cite{olatunji-etal-2023-afrispeech}     & \No  & \No  & \Yes & Mixed      & 1  & $\approx$123 & --  & --       & Afr.\ English   & ASR(WER) \\
SpokenWOZ\cite{si2023spokenwoz}                        & \Yes & \No  & \No  & Real-world & 2  & 249      & 5{,}700 & 203{,}000 & English        & ASR(WER), Task \\
VoxDialogue\cite{cheng2025voxdialogue}                 & \Yes & \No  & \No  & Mixed      & 2  & 42.56    & 4{,}500 & 30{,}700  & English        & LLM-Eval \\
MTalk-Bench\cite{du2025mtalkbench}                     & \Yes & \No  & \No  & Mixed      & 2  & 2.45     & 90      & 568      & English        & S2S, LLM-Eval, Human-Eval \\
\textbf{MMedFD (ours)}                                 & \Yes & \Yes & \Yes & Real-world & 2  & 136.91   & 1{,}805 & 10{,}814  & Chinese        & ASR(WER/HC-WER), LLM-Eval \\
\bottomrule
\end{tabular}

\begin{tablenotes}[flushleft,para]
\scriptsize
\textbf{Notes:} The symbols \Yes{} and \No{} denote presence and absence, respectively, ``--'' indicates information not reported. Definitions: \emph{Multi-turn} indicates explicit turn structure with cross-turn context, \emph{Full-duplex} refers to user--agent settings in which the system speaks while continuously listening, \emph{Healthcare} marks corpora collected in clinical/medical contexts. \emph{Nature} categorizes collection modality as \emph{Real-world} (naturally occurring audio), \emph{Simulated} (scripted/acted clinical dialogues), \emph{Read} (read speech), or \emph{Mixed} (combined sources). \emph{Scale}: \emph{Dur.\,(h)} denotes total hours, \emph{\#Dialogs} counts conversational sessions, \emph{\#Turns} counts utterance-level turns, \emph{Roles} indicates the number of distinct speaker roles. \emph{Evaluation} abbreviations: \emph{ASR} (Automatic Speech Recognition), \emph{WER} (Word Error Rate), \emph{CER} (Character Error Rate), \emph{HC-WER} (Healthcare-content WER), \emph{Task} (objective task metrics such as dialogue-state accuracy, slot F1, task success), \emph{LLM-Eval} (Large Language Model--based evaluation), \emph{Human-Eval} (human evaluation), \emph{TTS} (Text-to-Speech evaluation), and \emph{S2S} (Speech-to-Speech evaluation via arena/rubrics).
\end{tablenotes}

\end{threeparttable}
\end{adjustbox}
\end{table*}

\begin{table*}[!htbp]
  \centering
  \small % 保持 small 字号，在全宽下阅读体验最佳
  \begin{threeparttable}
    \caption{Role-wise ASR metrics across different Whisper variants on the train and test splits}
    \label{tab:role_metrics_models}

    % 适当增加行高，让宽表显得更大气
    \renewcommand{\arraystretch}{1.2}
    
    % 使用 tabular* 和 \textwidth 强制表格占满双栏总宽度
    % @{\extracolsep{\fill}} 会自动把剩余的空白均匀分配到列与列之间
    \begin{tabular*}{\textwidth}{@{\extracolsep{\fill}} ll cc|cc|cc @{}}
      \toprule
      \multirow{2}{*}{\textbf{Model}} & \multirow{2}{*}{\textbf{Role}} 
      & \multicolumn{2}{c}{\textbf{WER}} 
      & \multicolumn{2}{c}{\textbf{HC-WER (95\% CI)}} 
      & \multicolumn{2}{c}{\textbf{CER}} \\
      % 使用 \cmidrule(lr) 让子表头下方的横线在左右两端略微断开，层次更分明
      \cmidrule(lr){3-4} \cmidrule(lr){5-6} \cmidrule(lr){7-8}
      & & \textbf{Train} & \textbf{Test} 
      & \textbf{Train} & \textbf{Test} 
      & \textbf{Train} & \textbf{Test} \\
      \midrule
      
      % ================= whisper-small =================
      \multirow{3}{*}{\textbf{whisper-small}} 
      & \textbf{User}  & 54.56 & 53.11 & \valci{16.83}{15.51}{18.16} & \valci{15.37}{12.97}{17.77} & 50.51 & 52.08 \\
      & \textbf{Agent} & 1.92  & 1.84  & \valci{9.66}{9.17}{10.15}   & \valci{9.81}{9.22}{10.39}   & 1.86  & 1.81 \\
      & \textbf{All}   & 28.24 & 27.48 & \valci{13.25}{12.34}{14.16} & \valci{12.59}{11.10}{14.08} & 26.19 & 26.95 \\
      \midrule
      
      % ================= whisper-medium =================
      \multirow{3}{*}{\textbf{whisper-medium}} 
      & \textbf{User}  & 33.77 & 35.25 & \valci{35.96}{33.50}{38.42} & \valci{26.32}{22.30}{30.34} & 33.25 & 33.99 \\
      & \textbf{Agent} & 11.12 & 13.04 & \valci{9.39}{8.90}{9.88}    & \valci{14.50}{13.50}{15.50} & 10.44 & 11.48 \\
      & \textbf{All}   & 22.45 & 24.15 & \valci{22.68}{21.20}{24.15} & \valci{20.41}{17.90}{22.92} & 21.85 & 22.74 \\
      \midrule

      % ================= whisper-large =================
      \multirow{3}{*}{\textbf{whisper-large}} 
      & \textbf{User}  & 13.90 & 12.91 & \valci{21.35}{19.35}{23.35} & \valci{36.84}{31.84}{41.84} & 13.43 & 11.98 \\
      & \textbf{Agent} & 6.55  & 7.86  & \valci{12.40}{11.60}{13.20} & \valci{19.85}{18.35}{21.35} & 5.80  & 6.71 \\
      & \textbf{All}   & 10.23 & 10.39 & \valci{16.88}{15.48}{18.28} & \valci{28.35}{25.10}{31.60} & 9.62  & 9.35 \\
      \bottomrule
    \end{tabular*}

    \begin{tablenotes}[flushleft]\small
      \item \textbf{Notes:} Values are percentage means; 95\% CIs are in parentheses. 
      \emph{All} is the unweighted average of User and Agent for each metric and split.
      WER = Word Error Rate; HC-WER = Healthcare-aware WER; CER = Character Error Rate.
    \end{tablenotes}
  \end{threeparttable}
\end{table*}

\begin{table*}[!htbp]
  \centering
  \small % 全宽表格，保持 small 字号非常舒适
  \begin{threeparttable}
    \caption[LLM-judged healthcare QA results]{LLM-judged results for healthcare queries using PairEval and G-Eval with a consistent GPT-5 judge.}
    \label{tab:stacked_paireval_geval}

    % 增加行高，让表格呼吸感更好
    \renewcommand{\arraystretch}{1.2}

    % 展平为 7 列，并使用 \extracolsep{\fill} 强制占满整个页面宽度 \textwidth
    \begin{tabular*}{\textwidth}{@{\extracolsep{\fill}} l ccc|ccc @{}}
      \toprule
      \multirow{2}{*}{\textbf{Model}} 
      & \multicolumn{3}{c}{\textbf{PairEval (Pairwise Judge)}} 
      & \multicolumn{3}{c}{\textbf{G-Eval (Reference-Based Rubric)}} \\
      % 用左右断开的横线区分两个主要的 evaluation metrics
      \cmidrule(lr){2-4} \cmidrule(lr){5-7} 
      & \textbf{Win \% (95\% CI)} & \textbf{Tie\%} & \textbf{Any-Fail\%} 
      & \textbf{Overall (95\% CI)} & \textbf{Correct\%} & \textbf{Safety\%} \\
      \midrule
      
      GPT-5\cite{openai_gpt5}                
      & \valci{51.8}{49.0}{54.5} & 96.3 & 94.4 
      & \valci{3.9}{3.7}{4.1} & 35.2 & 35.2 \\
      
      Claude-Opus-4.1\cite{anthropic_opus41} 
      & \valci{50.1}{47.3}{52.8} & 99.8 & 99.8 
      & \valci{4.1}{3.9}{4.2} & 38.4 & 38.4 \\
      
      Gemini-2.5-Pro\cite{google_gemini25flash}
      & \valci{52.3}{49.5}{55.0} & 95.0 & 92.2 
      & \valci{4.0}{3.8}{4.1} & 36.4 & 36.4 \\
      
      Qwen3-30B-A3B\cite{qwen3_next80b}      
      & \valci{51.3}{48.6}{54.1} & 97.0 & 93.8 
      & \valci{4.1}{3.9}{4.2} & 37.7 & 37.7 \\
      
      Ours                                   
      & \valci{49.0}{46.2}{51.7} & 97.0 & 95.0 
      & \valci{4.0}{3.9}{4.1} & 36.9 & 36.9 \\
      
      \bottomrule
    \end{tabular*}

    \begin{tablenotes}[flushleft]\small
    \item \textbf{Notes:} PairEval compares two model outputs for each query using a fixed LLM judge; Win\% treats a win as 1 and a tie as 0.5, Tie\% is the share of ties, and Any-Fail\% is the share of responses that show any failure (repetition, off-topic content, role confusion, or contradiction). G-Eval scores each output against references on a 1--5 rubric; we report the mean and 95\% confidence intervals for \emph{Overall} (usefulness and coherence), \emph{Correct\%} (alignment with references and medical facts), and \emph{Safety\%} (avoiding harmful or inappropriate advice).
    \end{tablenotes}

  \end{threeparttable}
\end{table*}

\section{Discussion}
\label{sec:discussion}
Experimental results on MMedFD expose critical deployment bottlenecks in real-world full-duplex healthcare spoken dialogues. First, we observe a distinct trade-off between model scale and clinical precision: while scaling from \textit{whisper-small} to \textit{whisper-largev3} substantially improves overall User Test WER from 53.11\% to 12.91\%, it inadvertently inflates the User Test HC-WER from 15.37\% to 36.84\%, suggesting that larger foundation models under constrained domain adaptation tend to substitute low-frequency medical terms with phonetically similar general words. Second, the significant performance disparity between Agent and User subsets demonstrates that patient-side audio dominates full-duplex acoustic errors. This vulnerability is primarily driven by unconstrained reverberant environments, spontaneous prosody, and persistent overlapping speech during barge-ins, which severely disrupt streaming voice activity detection and speaker diarization sub-modules. Finally, downstream generation quality exhibits marginal variance across text-only LLMs when conditioned on identical upstream transcripts, with PairEval tie rates dominating between 95.0\% and 99.8\%. This performance equivalence indicates that response generation is primarily bounded by upstream acoustic processing rather than text-only reasoning capacities, strongly motivating a paradigm shift toward audio-native systems that jointly optimize acoustics, diarization, and context. Furthermore, mitigating these acoustic vulnerabilities is paramount for preserving multi-turn interaction, as cumulative upstream transcription errors can rapidly propagate and degrade downstream context tracking. Consequently, developing robust streaming error-recovery mechanisms and confidence-guided correction loops represents an essential prerequisite for deploying reliable, autonomous AI assistants in high-stakes clinical workflows.

\section{Conclusion}
\label{sec:conclusion}
We introduce MMedFD, a reproducible benchmark for multi-turn full-duplex Chinese healthcare dialogue. MMedFD integrates governance-compliant data collection, synchronized conversation view and user view audio, RTTM and CTM timing, and role labels, together with healthcare-grounded metrics and LLM-based evaluation. Experiments show that the primary bottleneck in automated clinical consultations is upstream acoustic processing. Current foundation models do not reliably trade off acoustic robustness and clinical terminology precision under spontaneous User speech with overlapping speech. By exposing errors in barge-in handling, role attribution, and clinical concept recognition, MMedFD provides a realistic testbed for streaming healthcare dialogue systems.

% \vfill\pagebreak

% \section{REFERENCES}
% \label{sec:refs}

\bibliographystyle{IEEEbib}
\bibliography{strings,refs}

\end{document}